\documentstyle[twoside,fleqn,espcrc2,epsf]{article}

\def\NPA{{Nucl. Phys.} {\bf A}}
\def\NPB{{Nucl. Phys.} {\bf B}}

\def\PRD{{Phys. Rev.} D}

\def\itmb{\begin{itemize}}
\def\itme{\end{itemize}}
\def\enmb{\begin{enumerate}}
\def\enme{\end{enumerate}}
\def\eqnb{\begin{equation}}
\def\eqne{\end{equation}}
\def\eqab{\begin{eqnarray}}
\def\eqae{\end{eqnarray}}

\newcommand{\AmS}{{\protect\the\textfont2
  A\kern-.1667em\lower.5ex\hbox{M}\kern-.125emS}}

\title{ Infrared Features of the Lattice Landau Gauge QCD}

\author{Hideo Nakajima\thanks{e-mail nakajima@is.utsunomiya-u.ac.jp}\\ 
Department of Information Science, Utsunomiya University, 321-8585 Japan \\
Sadataka Furui \thanks{e-mail furui@umb.teikyo-u.ac.jp}\\
School of Science and Engineering, Teikyo University, 320-8551 Japan}

\begin{document}

\begin{abstract}
The infrared properties of lattice Landau gauge QCD of $SU(3)$ are
studied by measuring gluon propagator, ghost propagator, QCD running coupling
and Kugo-Ojima parameter of $\beta=6.0, 16^4,24^4,32^4$ and $\beta=6.4, 32^4,48^4$ lattices. 
We study Gribov copy problem by parallel tempering (PT) gauge 
fixing of $\beta=2.2, 16^4$ $SU(2)$ configurations.
Rather unexpected features of the unique gauge(fundamental modular gauge, FMG) 
are observed in relation of physical values vs the optimizing function.
\vspace{1pc}
\end{abstract}

\maketitle

\section{The gluon propagator and the ghost propagator}

\begin{figure}[htb]
\begin{center}
\epsfysize=100pt\epsfbox{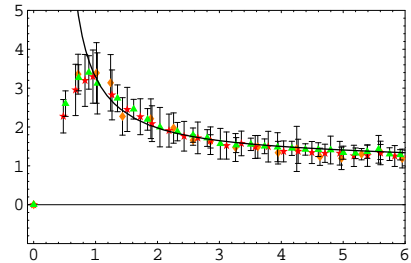}
\label{gl243248}
\end{center}
\caption{The gluon dressing function as the function of the momentum $q(GeV)$. 
$\beta=6.0$, $24^4$(triangle), $32^4$(diamond) and $\beta=6.4$, $48^4$(star) in $\log U$ version. 
The fitted line is $Z(q^2,y)$  in $\widetilde{MOM}$ scheme.}
\end{figure}

We measured $SU(3)$ gluon propagator of $\beta=6.0$ on lattice $16^4$, $24^4$, $32^4$
and $\beta=6.4$ on $32^4$, $48^4$\cite{NF,FN}. Shown in Fig.\ref{gl243248} is the gluon dressing function $Z(q^2)$ which is defined by the gluon propagator $D_{\mu\nu}(q)$ through
\eqnb
D_{\mu\nu}(q)
=(\delta_{\mu\nu}-{q_\mu q_\nu\over q^2})D_A(q^2)
\eqne
as $Z(q^2)=q^2 D_A(q^2)$. Here, we adopted 
the $\log U$ type definition of gauge field $A$. 
Although the size of the sample is not large, 
the data scale with each other and consistent with the result 
of\cite{adelade}. The fitted line is obtained from the model based on the 
principle of minimal sensitivity(PMS)\cite{vAc,PMS}. The PMS allows one 
to obtain a solution that bridges infrared region and ultraviolet region, 
but its uniqueness is not guaranteed and in the infrared region, 
there is discrepancy from other scheme that satisfies the commensurate scale 
relation\cite{BroLu}. We leave the continuation to the infrared region to 
the future study and fix the parameter $y$ by the solution at the 
factorization scale\cite{Gru,Orsay} i.e. $q=1.97GeV$ and call it the 
$\widetilde{MOM}$ scheme. 

The ghost propagator is defined as the Fourier transform of an expectation 
value of the inverse Faddeev-Popov operator $\cal  M$
\begin{equation}
D_G^{ab}(x,y)=\langle {\rm tr} \langle \lambda^a x|({\cal  M}[U])^{-1}|
\lambda^b y\rangle \rangle
\end{equation}
The lattice data scales and the PMS in $\widetilde{MOM}$ scheme fits in the 
region $q>0.5GeV$.
\begin{figure}[htb]
\begin{center}
\epsfysize=100pt\epsfbox{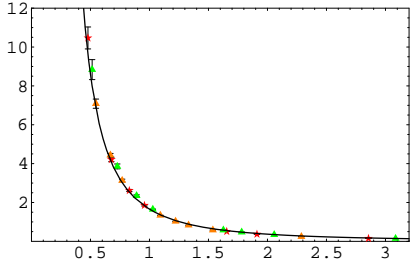}
\label{gh1}
\end{center}
\caption{ The ghost propagator as the function of the momentum $q(GeV)$. 
$\beta=6.0$, $24^4$(triangle), $32^4$(diamond) and $\beta=6.4$, $48^4$(star) in $\log U$ version. 
The fitted line is $Z_g(q^2,y)$ of the $\widetilde{MOM}$ scheme, which is 
singular at $q\simeq 0.4GeV$.}
\end{figure}

\section{QCD running coupling and Kugo-Ojima parameter}

In the contour-improved perturbation theory\cite{HoMa}, physical quantities 
${\cal R}$ are expressed in a series
\begin{equation}
{\cal R}(q^2)={\cal B}_1(q^2)+\sum_{n=1}^\infty A_n{\cal B}_{n+1}(q^2)
\end{equation}
where $A_n$ is known by perturbation theory in loop expansion, and 
${\cal B}_n$ are expressed by the Lambert W function.
The result of running coupling using $\Lambda_{\overline{MS}}=237MeV$ fits the data of 
$\beta=6.4, 48^4$ data, except the peak around $1GeV$.  
\begin{figure}[htb]
\begin{center}
\epsfysize=100pt\epsfbox{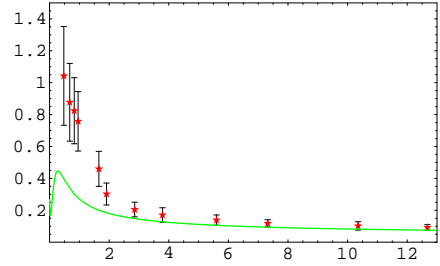}
\label{alp32}
\end{center}
\caption{ The running coupling $\alpha_s(q)$ as a function of  
momentum $q(GeV)$. $\beta=6.4, 48^4$ lattice. Fitted line is the result
of the contour improved perturbation method.}
\end{figure}

In the $\beta=6.4, 48^4$ lattice in $\log U$ version, we obtained the 
Kugo-Ojima parameter $c=0.793(61)$\cite{KO}, trace $e/d=0.982(1)$ and 
the horizon function deviation $h=-0.19$\cite{Zw,NF,FN}.

\section{Gauge fixing in use of PT and Gribov copy problem in $SU(2)$}
Fundamental modular gauge (FMG: global maximum of the optimizing 
function $F_U(g)$)\cite{Zw} is the unique gauge known as the legitimate Landau gauge
without Gribov copies\cite{Gr}, where $F_U(g)=\sum_{\ell}{\rm Re\ tr}U_\ell^g$. 
Simulated annealing methods are now widely used to cope with Gribov copies in 
the maximal abelian gauge\cite{BBMS}. In the present work, we combine 
parallel tempering (PT) method 
as a new trial with gauge fixing for FMG, and test it in $SU(2)$ gauge theory
on $16^4$ lattice with $\beta=2.2$ of the Wilson action. The PT was invented as 
one of suitable methods for 
simulating correctly low temperature equilibrium of systems possessing 
complex energy profile like spin glass\cite{HN,MKH}. We do not need the exact
equilibrium of Boltzmann distribution ${\rm Prob}(g)\propto exp\{\beta F_U(g)\}$, 
but only the lowest energy state $g$ of $Min_g E(g)$ efficiently 
where $E(g)=-F_U(g)$. 
In the PT method, we prepare $N_r$ replicas of site-spin (gauge transformation) 
system with gauge link interaction $U$. 
Each replica $g_j$ is given different $\beta_{i_j}$, 
$\beta_{min}\le \beta_i\le \beta_{max}$. 
After $N_{mc}$ Monte Carlo steps at the given $\beta$, 
the $\beta$'s are exchanged between adjacent values according to
Metropolis acceptance check\cite{HN,MKH}. 
During $N_{ex}$ iterations of the 
$\beta$-exchange step, the lowest energy state among all replicas is
monitored and stored. These accumulated $N_c$ gauge spins are 
all fixed to the Landau gauge by the standard over-relaxation (OR) method 
(${\rm Max}_x(\partial A)_x<10^{-4}$). 
We pick up the best copy among those, and name it as PT copy. 
In the PT method,
the distribution of $\beta$'s in the interval $[\beta_{min},\beta_{\max}]$ 
is technically the most important point so that the exchange rates 
are almost equal everywhere between replicas.
We found the good choice for adjustment of
the new $\beta'$'s from the old $\beta$'s as  
$\triangle\beta'\propto \sqrt{\bar{\Delta}/(\triangle E/\triangle\beta)}$ where 
$\Delta=\triangle\beta\cdot\triangle E$ and $\bar{\Delta}$ is the replica 
average of $\Delta$. Due to the actual situation of our choice, 
$\beta_{min}=0.4$ and $\beta_{max}=30$, and the system size of $16^4$, 
and $N_r=24$, direct application of the PT is practically meaningless 
since the Metropolis acceptance rate is very small. We dare to shrink
the energy scale by a factor $\alpha=0.0002$ only within the $\beta$ exchange 
procedure
to have the Metropolice acceptance rate around $0.5$. 
Other parameters of ours are as follows, $N_{mc}=120$, $N_{ex}=12$, 
$N_c=16\sim 32$. Total number of PT gauge fixed samples of $\beta=2.2$ 
of $SU(2)$ Wilson action is 67, and we compared them with gauge copies
that are directly gauge fixed by OR method and are named as first copies. 
Among 67 samples, 63 PT copies give larger $F_U(1)$ with $N_c=16$ 
than first copies and the remaining 4 PT copies with $N_c=20\sim 32$.
Using these two groups of copies, Gribov noise is investigated in
Kugo-Ojima parameter, gluon propagator and ghost propagator.
The averaged Kugo-Ojima parameter is suppressed about 4\% in PT in comparison
with first copy. Effects of noise about 6\% are seen in Fig.4 for the ghost propagator, but no significant effects in the gluon propagator. 
For a few samples among 67, comparison is made for each sample 
between two groups of $16(=N_c)$ PT generated copies, 
and 32 copies generated(CRGT) by direct OR gauge fixing
after random gauge transformation, in a scatter plot
of Kugo-Ojima parameter vs $F_U(1)$.
It is observed in this plot that a group of PT copies scatter almost in a line on top-end of $F_U(1)$, with all CRGT scattering in lower $F_U(1)$ area, and this
fact raises questions if this "flat-end" feature is real one of
the FMG, or if we have not reached the real FMG yet? This is a rather unexpected
feature with respect to FMG. 

\begin{figure}[htb]
\begin{center}
\epsfysize=100pt\epsfbox{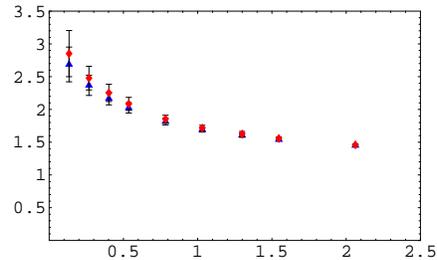}
\label{gh67}
\end{center}
\caption{The $SU(2)$ colour diagonal ghost dressing function as the function of the lattice momentum squared $q^2a^2$. 1st copies(diamond) are more singular than PT(triangle).}
\end{figure}

\section{Discussion and outlook}
The momentum dependences of the gluon propagator and the ghost propagator
in the infrared are analyzed in use of PMS method.
Deviation of $u(0)=-c$ from -1 (Kugo-Ojima confinement criterion) does not necessarily imply 
violation of colour symmetry. It could imply a question on the realization 
of BRST quartet mechanism on the lattice and/or even further validity of 
the continuum Lagrangian in the infrared region. So far as our data in
off-diagonal elements of ghost propagator show, 
there appear no definite signals of colour global symmetry breaking.

From the scatter plot the Kugo-Ojima vs $F_U(1)$, it seems that our PT gauge fixing
works well, but more careful parameter tuning and comparison with 
simulated annealing method deserve a future study.
Gribov noise in the ghost propagator is qualitatively similar to 
Cucchieri's noise at smaller $\beta=1.6$\cite{Cch}. The infrared singularity is weakly
suppressed in FMG.

Some revisions in numerical data of ghost and Kugo-Ojima parameter have been made in this paper due to use of conjugate gradient method.

This work was supported by the KEK supercomputing project No.03-94.

\end{document}